# Gentlemen on the Road:
# Understanding How Pedestrians Interpret Yielding Behavior of Autonomous Vehicles using Machine Learning.

**Yoon Kyung Lee**[1], **Yong-Eun Rhee**[1], **Jeh-Kwang Ryu**[2], and **Sowon Hahn**[1]

[1]Department of Psychology, Seoul National University,
[2]Department of Physical Education, Dongguk University
{yoonlee78, clara.rhee, swhahn}@snu.ac.kr, ryujk@dongguk.edu

**Abstract**

Autonomous vehicles (AVs) can prevent collisions by understanding pedestrians' intention. We conducted a virtual reality experiment with 39 participants and measured crossing times (seconds) and head orientation (yaw degrees). We manipulated AV yielding behavior (no-yield, slow-yield, and fast-yield) and the AV size (small, medium, and large). Using machine learning approach, we classified head orientation change of pedestrians by time into 6 clusters of patterns. Results indicate that pedestrians' head orientation change was influenced by AV yielding behavior as well as the size of the AV. Participants fixated on the front most of the time even when the car approached near. Participants changed head orientation most frequently when a large size AV did not yield (no-yield). In post-experiment interviews, participants reported that yielding behavior and size affected their decision to cross and perceived safety. For autonomous vehicles to be perceived more safe and trustful, vehicle-specific factors such as size and yielding behavior should be considered in the designing process.

# 1 Introduction

Autonomous driving is an ever-fast developing technology world-wide. While recent successful cases of autonomous cars on the public roads signal exciting news, it raises safety concerns at the same time. Korea is one of the leading countries in developing autonomous driving technology, yet it has the highest pedestrian fatality rates. The number of deaths in traffic accidents per 100,000 populations in Korea was 4.1 in 2014, which is three times the OECD average (1.4). In the same year Korea had the highest rate of pedestrian fatalities anywhere in the OECD and the number of elderly fatalities per capita was triple the OECD average (Adler and Ahrend, 2017). However, letting autonomous vehicles drive on the real road can lead to multiple challenges not only in the context of driver-vehicle interface, but also in pedestrian-vehicle communication. In addition, social acceptance of AV and public trust should precede operating AVs on public roads. Researchers have shown more interests on Human-AV interactions recently (Ackermann et al., 2019; De Clercq et al., 2019; Dey and Terken, 2017). However, much of this research focused on looking at pedestrian reaction from a driver's perspective. There is a lack of studies focusing on the pedestrian perspectives. It is necessary to examine communication means people use in the face of vehicles on the road.

# 2 Related Works

## 2.1 Trust in AV

When an artificial agent sends a social signal (e.g., greeting), our expectation that it will continuously show more human-like social response increases (Reeves and Nass, 1996). But when the agent's behavior is hard to interpret (e.g., "Will that driver stop in front of me or not?"), our trust in the agent decreases. Studies show that the more uncertain the vehicles' intentions are, the less the pedestrians trust them (Jayaraman et al., 2019). Pedestrians showing trust in the vehicle's intention to ensure their safety generated reduced crossing speed, less frequent staring at vehicles, and shorter distances to collision at signalized crosswalks (Asaithambi et al., 2016; Rasouli et al., 2017; Tom and Granié, 2011). In unsignalized crosswalks, the rules may not be clear which of the cars and pedestrians should go first. In signalized crosswalks, people rarely engage in interacting with vehicles because both parties are expected to follow the traffic signals. We wanted to see how pedestrians interact with the AVs when there is no clear rule of who goes first. We focused on interactions in unsignalized crosswalks. We expect that pedestrians will engage more in communicating with the approaching vehicle in an unsignalized crosswalk than in a signalized crosswalk.

## 2.2 Movement as Intent Communication

Movements and gestures are important to coordination and performance of joint activities with which they communicate intentions. When given dot points moving, human can derive one's own interpretation of their purpose, cause, and expected results (Dittrich et al., 1996; Heider and Simmel, 1944). Movement is used as a way to express socially appropriate behavior or to communicate intent to create social distance (Ekman and Friesen, 1969). Autonomous robots also use movement as a communication method (Lehmann et al., 2015; Thompson et al., 2011). Social signals delivered via movements such as nodding, shaking hands, advancing or retreating from others can be interpreted in different ways depending on the situations. Teaching machines to behave in socially appropriate ways can make cooperation with humans smoother in several cases like manufacturing, sports, and even a jazz ensemble (Hoffman and Ju, 2012). Recently, there have been increasing cases of research on AV's socially appropriate movement gesture as effective non-verbal communication means affecting pedestrian trust (Asaithambi et al., 2016; Risto et al., 2017). In crosswalks, it is difficult to verbally communicate, so both pedestrians and drivers put efforts on interpreting each agent's movements. Studies show that pedestrians use many nonverbal communication means such as raising hands, staring, race walking, and bowing (Guéguen et al., 2015; Rasouli et al., 2017; Ren et al., 2016). These messages can be also understood differently depending on the culture and locations. Pedestrians assume that the AV will ensure the safe distance between them, and if this social agreement is violated, they react accordingly. (e.g., staring, putting up a hand).

## 2.3 Intent Communication of AV: Yielding Behavior

Autonomous vehicles (AVs) need the ability to communicate their intent with pedestrians. Previous studies on human-AV interaction, however, have mainly concerned communication between the driver and AV (Bellem et al., 2018; Buckley et al., 2018; Lee et al., 2019; Seppelt and Lee, 2019). It is crucial for the related fields to study how pedestrians perceive safety and use both verbal and non-verbal communication means to safely cross. Studies show that AVs that have an interface showing its intent of waiting or passing were perceived safer than those without it (Böckle et al., 2017; Habibovic et al., 2018).

Keeping a safe distance and showing the desire and intention of doing so will be regarded as socially appropriate manners. If AVs behave like how good-mannered human drivers would yield, this behavior could increase the trust and perceived safety in pedestrians as well. To make AVs socially appropriate, we should regard vehicles as social entities that can affect the behaviors and psychological states of pedestrians. A vehicle approaching without slowing the speed invades the comfort boundaries and thus evokes pedestrians' emotional responses such as fear (e.g., stopping or running) and discomfort (e.g., staring). Many pedestrian responses to drivers rely on subtle and non-verbal cues, which sometimes lead to miscommunication or perceived risks.

However, only a limited number of studies investigated pedestrian-AV interaction. Recent approach on studying pedestrian-AV interaction are on-site observation of pedestrian reaction to "driverless" vehicles, which were driven by human drivers hidden behind a car seat (Dey et al., 2019; Palmeiro et al., 2018; Risto et al., 2017; Rothenbücher et al., 2016). Previous studies mainly focused on vehicle factors that affected pedestrian behaviors such as vehicle appearances, speed, and presence of crosswalks (Rasouli and Tsotsos, 2019; Schmidt and Faerber, 2009; Sucha et al., 2017). Recent studies on pedestrian-AV interactions focused on manipulating Head Mount

Displays (HMD) in order for vehicle to deliver its intent (e.g., yielding, passing; Habibovic et al., 2018; Mahadevan et al., 2018). However, HMDs are useful only when pedestrians are always watching the front of the vehicle. It is known that pedestrians rarely look at the human-driven vehicles until it "misbehaved" by approaching aggressively and not slowing its speed (Rothenbücher et al., 2016). Pedestrians showed a blind trust in the driver and did not notice until the vehicle advanced and did not yield. In other field study by Risto et al., 2017), pedestrians showed discomfort by staring at vehicles when they did not keep a safe enough distance between them. We on the other hand will focus on the vehicle movement, yielding. And to make it more human-like, we will manipulate the speed of yielding behavior so that the AV 'intends' to give a sufficient safe distance from the pedestrian.

### 2.4 Intent Communication of AV: Yielding Behavior

Pedestrians were mostly viewed as moving articles in many past studies. Therefore, motion itself has served as a strong parameter of the pedestrian movement and safety. Commonly observed trusting behaviors of both drivers and pedestrians include lack of monitoring the AVs (e.g., low gaze ratio and head movement (Hergeth et al., 2017; Jayaraman et al., 2019). For pedestrians in specific, this lack of monitoring may lead to risky behaviors such as jaywalking and allowing close distances to the vehicles, which are the instances of *overtrust* (Parasuraman et al., 1993).

Recent approach to pedestrian intent communication and estimation has suggested looking at non-verbal parameters such as hand wave, eye contact, and verbal expression. Head orientation has served as an important index for pedestrian crossing intent (Kooij et al., 2019; Kwak et al., 2017; Rasouli et al., 2017; Rasouli and Tsotsos, 2019; Schulz and Stiefelhagen, 2015). Here we suggest using head orientation of a pedestrian as an indicator of cautionary behavior - hereinafter expressed as 'looking around behavior' measured by 'head orientation'. Individual differences in head orientation is expected as we as pedestrians do not share identical behavioral patterns while crossing: some people do not look at the vehicles at all whereas someone others are more careful about crossing. Using machine-learning based clustering classification methods, we expect to identify several different types of cautionary behaviors in Korean pedestrians.

### 2.5 The Present Study

We used a virtual environment to simulate a road condition that is similar to the typical Korean one-lane road. Using virtual reality, one can measure various factors that can affect pedestrian behavior. Recent bodies of research use virtual reality as an alternative to the pre-existing methods. Studies on crossing safety education, and human-robot collaboration used virtual reality (Matsas et al., 2018; McComas et al., 2002; Whitney et al., 2018; Zanbaka et al., 2007). In virtual reality, people also exhibit a natural response to the virtual character/agent so we expect that pedestrians in our virtual crosswalk setting will do so. To ensure this, we will also conduct a user-test to check whether the environment itself did not hinder the VR experience.

Research shows that simple addition of a factor in a dynamic traffic situation can lead to catastrophic consequences (Rasouli and Tsotsos, 2019). Hence, it will be difficult to experiment in simulated roads with real vehicles. Using virtual reality, one can derive both motion and non-motion reactions of pedestrians. Virtual reality can safely reproduce different countries, road sizes, and weather, allowing you to test factors that affect pedestrian safety. Simply adding realistic settings such as buildings or landmarks that exist in reality into virtual reality settings can increase the sense of realism and presence. As researchers, we can consider different environmental and contextual factors as manipulating variables such as country settings, volume of traffic, and even hazardous weather conditions.

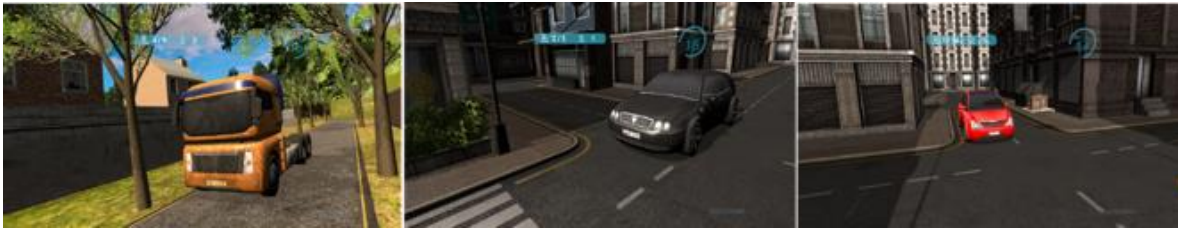

Figure 2. Autonomous Vehicle Stimuli Used in the Experiment: Large (left), medium (control condition; center), and small size AVs (right).

## 3 Methods

### 3.1 Participants

A total of 37 people participated in the study ($M_{age}$=24.14, $SD_{age}$=4.92, $N_{female}$=17). Only participants who reported no experience of dizziness, nausea, or vomiting following any virtual-reality related experiences were eligible to participate. The study was approved by the Institutional Review Board of Seoul National University (IRB No.1807/002-012) and carried out in accordance with the approval including all guidelines. Participants informed us of their consent after safety instructions were given.

### 3.2 Materials

We used a HTC Vive Pro head-mounted-display (HMD) with resolution of 2880 x 1660, a viewing angle of 110 degrees, and a frequency of 90 Hz. We set up a TPCast wireless VR adapter for wireless connection of VR equipment. Head tracking was achieved in real time. Participants were able to listen to the sound via the headphone equipped to the HMD. We set two Lighthouse sensors diagonally across the room to track head orientation. For safety reasons, 4 barricades were put at each corner of the room (Figure 1).

The virtual crosswalk task (VCT) was executed on a desktop computer with an Intel Core i7-7700CPU, with 16GB of RAM and an NVIDIA GeForce GTX 1070, running Windows 10 and DirectX 11. We used Unity 2017 3D software to simulate the typical look of downtown and rural roads in Korea. We used the Steam VR library to design components such as buildings, landscapes, and crosswalks.

### 3.3 Design

**Yielding Behavior.** AV's yielding behavior was defined as waiting for pedestrians to cross followed by an explicit deceleration. Different yielding speed (slow-yield and fast-yield) was manipulated by the speed of deceleration. AVs in fast-yield condition reduced the speed immediately after detecting the pedestrian and completely stopped from a longer distance. AVs in slow-yield condition have done the same as in fast-yield condition, except the decelerated slowly and stopped. In a shorter distance. We set. The initial speed of vehicles in a range of 3.33 m/s, 5.56 m/s, 4.17 m/s to prevent participants from easily noticing the pattern of the AV. When participants were detected, all AVs slowed down to 1.11 m/s. In slow-yield and fast-yield trials, AVs stopped when participants were within a close distance of 0.5m. If no participant were present, AVs proceeded without yielding. Prior to the experiment, we conducted a user-test with psychology experts (N = 4, 2 females, Mage = 27) to find the right speed, distance, and brake power for AVs in different conditions. We finalized the setting as follows: all AVs had a maximum speed of 19 km / h and an acceleration of 0.80 km / h. The recognition distance was set at 4.5 m for fast-yield and 7.0 m for both no-yield and slow-yield. In fast-yield condition, brake power was set to 6.0, and 3.0 for AVs in both no-yield and slow-yield conditions. In slow-yield, AVs stopped for 2 seconds in order to detect the presence of participants within the given range of 0.5 m. In no-yield condition, all AVs passed slowly in crosswalk with a speed of 4.0 m/s.

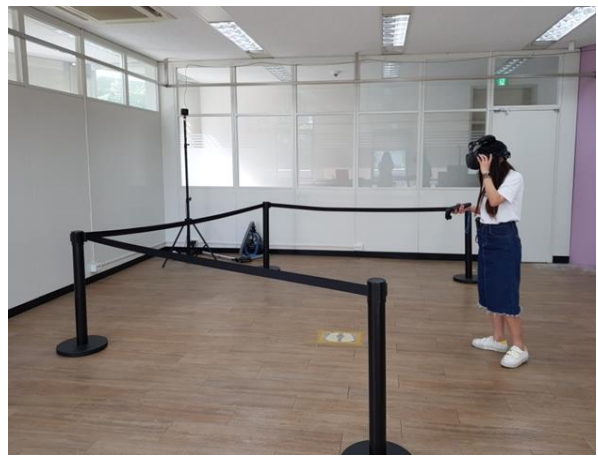

Figure 1. Virtual Reality Experiment Setup

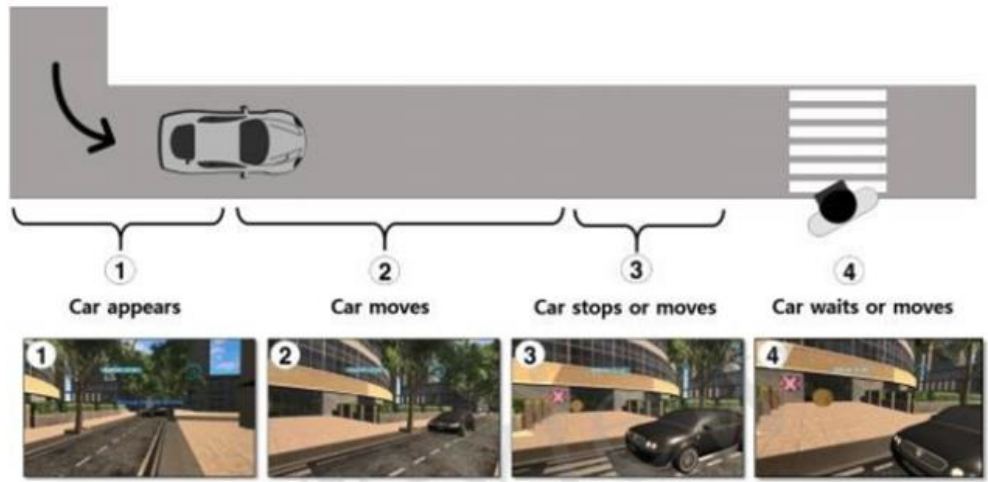

Figure 3. Virtual Crosswalk Experiment Procedure

**AV Size.** There were 3 different sizes of AVs used: small, medium, and large. AVs were replications of typical sizes of compact AVs, medium sedans or vans, and trucks, respectively (Figure 2).

### 3.4 Procedure

Participants were first informed of the general procedure and safety instructions regarding the experimental room setup. At each corner of the experimental room, we installed four barricades. As participants put the HMD on, researchers adjusted the amount of pressure in order for a proper fit. Two researchers were assigned to each experiment: one monitored the HMD connection and observed what participants viewed and the other safely guarded participants. We also instructed participants to walk as if he/she were walking on a real crosswalk and not to get hit by the AV. Participants were randomly assigned to at least 2 instances of 9 conditions: 3 yielding behavior (no-yield (control), slow-yield, and fast-yield) x 3 AV size (small, medium(control), large). We conducted a total of 3 practice trials with no companion, small size AVs, and city downtown setting of virtual crosswalk. In test trials, participants crossed zebra crossings that were located in either one of three settings: 2 urban and a rural one-lane roads.

Figure 3 shows a general procedure of a single trial in the virtual crosswalk. When the trial started, a "ready" sign appeared and 3 seconds of waiting time was given. After the waiting time ended, the trial started and a time-counter indicating the time limit for each trial started. Time limit was 20 seconds for each trial. At the start of the trial an instruction saying "cross the road and receive a coin at the end of the crosswalk" was shown. We put a time-counter and a coin-based reward system in order to keep the experiment interesting and not to subject participants to experience boredom quickly. At the end of the crosswalk, there was an arrow indicating the end-point. If the participants succeeded at the trial, the message "success" was shown and a button pressed by the participant set up the next trial. Participants then had to turn around in order to start the new trial. A new arrow then appeared either at the same location that the participant just arrived at or at the opposite side so that a new crosswalk is made throughout the room in diagonal direction. Within 200 ms, an AV started to appear and the participant could hear the sound of the AV initiating the engine. The AV then turned around the corner and started approaching the crosswalk. Then, when the distance between the AV and the participant was 0.5 m, the AV stopped in slow-yield and fast-yield conditions. The AV did not stop at all in no-yield condition trials. The location of the vehicle was randomized to either the left or right side of the participant. Throughout the

|  | Yielding Behavior | | |
|---|---|---|---|
|  | No-Yield* | Slow-Yield | Fast-Yield |
| Size | M(SD) | M(SD) | M(SD) |
| Large | 10.63 (5.65) | 9.70(5.08) | 8.54(3.55) |
| Medium* | 9.61(3.69) | 9.14(2.80) | 9.04(3.48) |
| Small | 10(4.59) | 8.92(3.73) | 8.38(3.57) |

Table1. Mean Crossing Time of Pedestrians (Seconds). *Control conditions

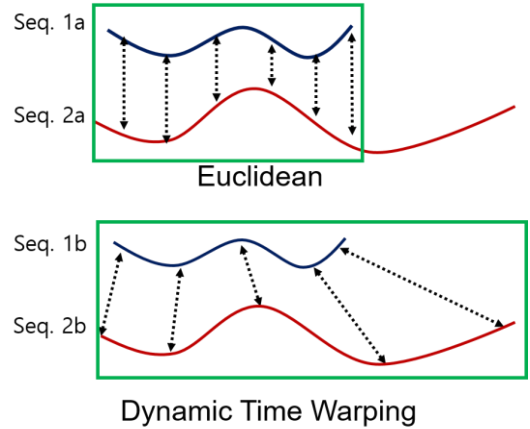

Figure 4. Comparison of Time Series Pattern Similarity Computation Using Euclidean Distance and Dynamic Time Warping. Sequence 1a and 2a, 1b and 2b are two patterns found in time series data and differ in time length. Green box indicates the window of time series that are being compared. Dynamic Time Warping is advantageous over using Euclidean distance as it can stretch out to find similar patterns between the two-time series, hence resulting in more chance to find precise similar patterns.

experiment, we measured the total duration of crossing time, success rate, and head orientation (0 to 360 degrees) by 200 ms which were measured automatically by HMD. When finished, we conducted post-experiment interviews.

We asked participants to give their opinions about the overall experience. The questions were devised by the authors. We collected Realism ("how real did you feel of the virtual experience?"), Similarity to Real-World Crossing Behavior ("how similar did you cross the road to the real crossing behavior?"), Effect of AV Size ("how much did the AV's size affect your crossing behavior?"), and Yielding Behavior of AV ("how much did the AV's yielding behavior affect your crossing behavior?"). We also asked to describe in detail any attempt to communicate or take caution when faced with AVs in the experiment.

## 4 Results

### 4.1 Success Rate and Crossing Time

All 37 participants attempted to cross the crosswalk. No participant hesitated or declined to cross. However, not all participants were able to successfully cross without either getting hit by the AV ('fail-to-cross') or excess time limit ('timeout'). For analysis, we excluded 23 trials that were either 'fail-to-cross' or 'timeout' (0.02% of our sample). Average success rate was 98.7%. We also examined how long it took for participants to cross. We calculated the mean duration of crossing time (from the start point to the end point). Mean crossing times per condition are listed in Table 1. Participants crossed the crosswalk on average of 8.76 seconds (SD = 4.23 sec). In general, the shortest time was less than 1 second and the longest was 19.99 seconds. Participants took the longest time to cross when large AVs did not yield (M = 10.63, SD = 5.65), whereas they took shortest when AVs were small and yielded fast (M = 8.38, SD =3.57).

### 4.2 Head Orientation

Previous studies used head orientation to detect pedestrian intention to cross and trust (Rasouli et al., 2017; Kooji et al., 2014; Schulz & Stiefelhagen, 2015; Metaxas & Zhang, 2013). We classified patterns of head orientation change and examined the effect of AV's yielding behavior. Plus, we also looked at whether this effect is modulated by the size of the AV. We divided each time sequence into 4 periods to get a precise scope of behavioral change in reaction to AV's behavior. We annotated the events: AV appeared, approached, stopped, and waited (or passed). The visualization of each AV's action is illustrated in the screenshot at the bottom of Figure 2. AVs in no-yield condition passed instead of waiting. Each point of head position was polar coordinates $(x,y,z)$. We collected yaw degrees because we were interested in a behavioral index that is equivalent to a looking-around (left to right) behavior. The HMD has a limited field-of-view of 105 degrees. Due to the limited field-of-view, we defined 'front' direction that is different from real field of view in a natural setting (which can be up to 180 degree). In polar coordinate degree, we defined head drift between 90 and 270 degrees as front.

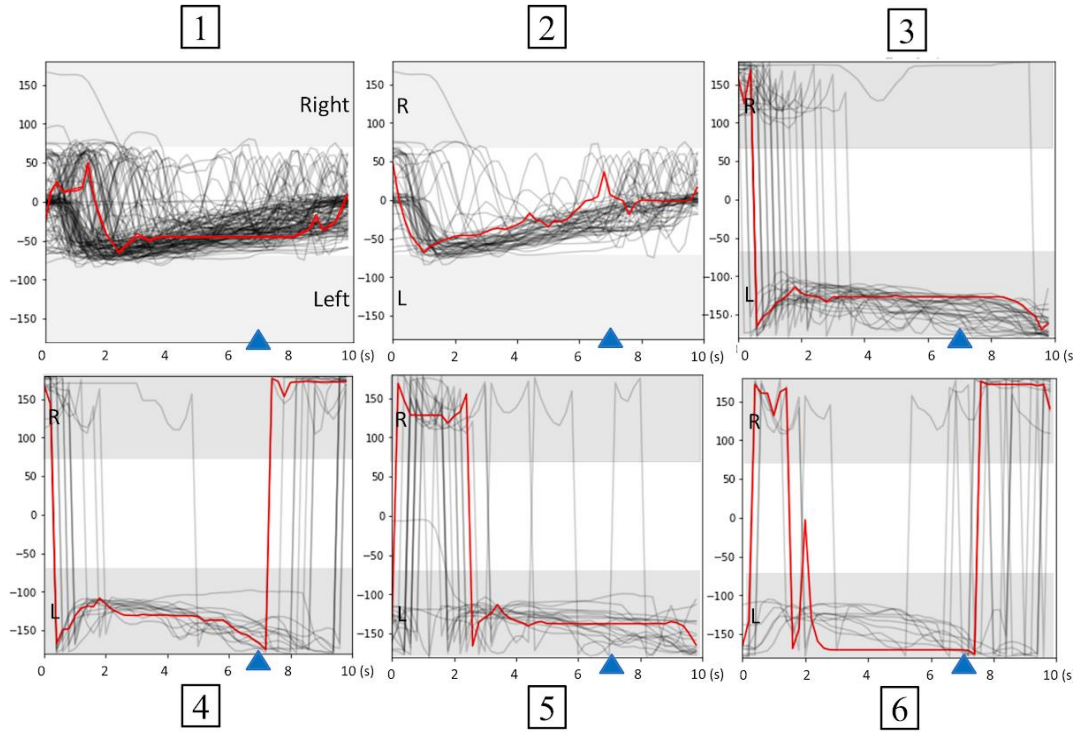

Figure 6. Illustration of Head Orientation Change Patterns. Red line refers to the feature time series pattern found by the DTW algorithm. In Cluster 1 and 2, participants only fixated on the 'front'. In Cluster 2, slightly shifting the head orientation from -55 degree to 0 degree can be found. In Cluster 3, participants fixated their head toward the opposite direction of a vehicle approaching with limited head movement. In Clusters 4 and 5, participants continued to see the opposite direction until the car stopped. In Cluster 5, participants only moved their head when the AV approached. In Cluster 6, pedestrians 1) looked toward the direction of an AV approaching, 2) spotted the vehicle, 3) looked towards the opposite side, 4) and then looked toward the AV

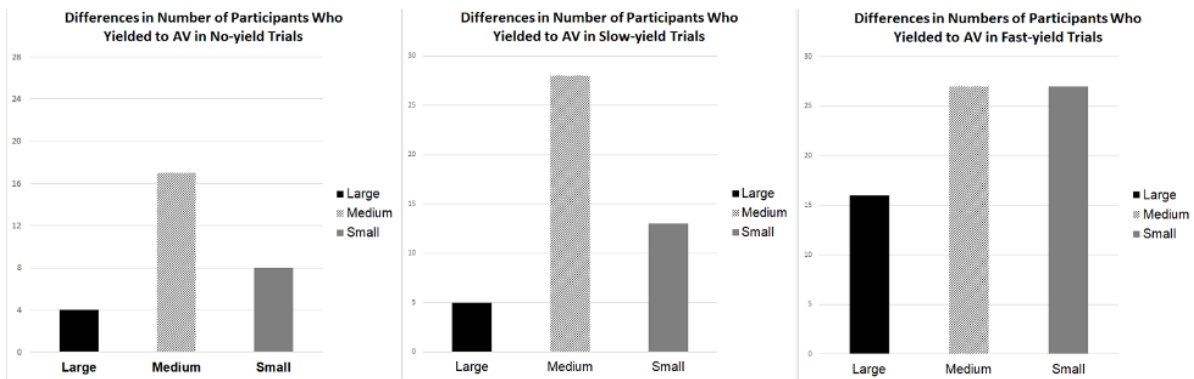

Figure 5. Numbers of Participants Who Yielded to AV. In no-yield and slow-yield trials, participants were less likely to yield in both large and small AVs compared to the medium-size (control condition). Overall, participants were more likely to yield to medium size, small size, then large size AVs. Participants were generally more likely to yield when AV yielded as well than when AV did not yield.

For clustering head orientation change patterns, we used Dynamic time warping (DTW), a time series pattern recognition algorithm that measures the optimal similarity between two time sequences that differ in time length. It is widely used in fields like speech recognition, futures trading in systems, and graphic or video pattern recognition. It matches the two time series in a direction that minimizes the distance between them. As shown in Figure 4, when two time series are matched using DTW, it can be appropriately matched to a set of waveform that are partially distorted, unlike when using the Euclidean distance method (Keogh et al., 2001). For our analysis, we used Soft-DTW (Cuturi and Blondel, 2017), a time-series pattern classification algorithm based on DTW which has shown better performance recently. We used DTW because our data vary in time length. We found great individual differences in crossing time (min. 1 sec. to max. 19.9 sec.). The angle of head orientation was collected in every 200 ms and the range was 0 to 360 degrees. We converted the yaw degree in order to get a more comprehensive visualization of the looking-around gesture indicated by head orientation change from left to right. Assuming 0 degree as front, we coded the head orientation to the far right as -180 degrees and 180 degrees as head orientations to the far left.

|  | **AV Behaviors** | | | |
| --- | --- | --- | --- | --- |
| Clusters | Appear | Approach | Stop* | Wait* |
| 1 | *Front* | *Front* | *Front* | *Front* |
| 2 | **Left** | **Left** | *Front* | *Front* |
| 3 | Right to left | **Left** | **Left** | **Left** |
| 4 | Right to left | **Left** | **Left** | <u>Right</u> |
| 5 | <u>Right</u> | Right to left | **Left** | <u>Right</u> |
| 6 | <u>Right</u> | <u>Right</u> | **Left** | **Left** |

Table 2. Description of Head Orientation Change in Reaction to AV Behaviors: Descriptions were based on agreement of experts (*n* = 3). Although the average line shows the head orientation was toward the left (opposite of the vehicle approaching), it was still within the range of 'front' band of yaw degrees (-55 to 55 degree). ***Left** = head direction against approaching vehicles, <u>Right</u> = head orientation toward approaching vehicles, *Front* = neither against nor toward approaching vehicles. Experts interpreted head direction to left as gesture of high level and right as the lowest level of trust in AV.

In each trial, it took 3.5 seconds on average for the AV to appear and start approaching. We observed that participants showed a wide range of different time and even the same participant showed greater difference in crossing time by trials.

|  | Cluster 1 | Cluster 2 | Cluster 3 | Cluster 4 | Cluster 5 | Cluster 6 |
| --- | --- | --- | --- | --- | --- | --- |
| SY-Large | **63%** | - | 20% | - | - | 17% |
| NY-Large | **57%** | - | - | 18% | 25% | - |
| FY-Large | - | **60%** | 8% | 16% | - | 16% |
| SY-Small | - | **52%** | 16% | - | 20% | 12% |
| FY-Small | - | 14% | 18% | - | **68%** | - |
| NY-Small | 23% | - | - | 29% | 10% | **39%** |

Table 1. Ratio of Clusters in AV Behavior by AV Size Conditions

In order to conduct time series classification, the time sequence shorter than 2 seconds and longer than 10 seconds were normalized into a total range of 10 seconds. A total of 248 sequences were analyzed.

Prior to classification, average time sequences with respect to $dtw_\gamma$ discrepancy were computed. Each $dtw\gamma(x,y_i)$ was divided by $m_i$, the length of $y_i$ in Equation 1. We used k-means clustering to extract commonly observed head orientation change patterns and the number of each pattern observed by conditions. K-means clustering is a machine learning based classification method for grouping data based on the similarity of average points. As formally stated in Equation 2, generalization of Loyld algorithm for k-means clustering (Lloyd, 1982) was conducted in each centering and clustering allocation step according to the DTW lambda discrepancy (Equation 3; See Cuturi and Blondel, 2017).

$$\min_{x_1,\dots x_k \in R^{p*n}} \sum_{i=1}^{N} \frac{1}{m_{ij}} min_{\in [[k]]} dtw_\gamma(x_i, y_i) \quad (1)$$

$$\min_{x \in R^{p*n}} \sum_{i=1}^{N} \frac{\lambda_i}{m_{ij}} dtw_\gamma(x_i, y_i) \quad (2)$$

$$k^* = argmin \sum_{i=1}^{k} \sum_{j \in S_i} |X_i - \mu_i|^2 \quad (3)$$

Figure 6 illustrates head orientation change by time of events of behaviors of AV. Table 2 shows a description of head orientation of participants on each time of the events. We normalized the time length to 10 seconds to compare head orientation change at the same time. X axis refers to seconds after the start of the trial. Y axis is the yaw degree of the head position that is converted into the range of -180 and 180. Maximum field of view was between -55 to 55 degrees when facing front. Yaw degrees with a range from 55 degrees to 180, were defined as the looking right (toward the AV), and -55 to -180 to the left (opposite of the AV). We trained the dataset into a time series classification using python (Tavenard et al., 2020).

We found a total of 6 clusters of head orientation change patterns (Figure 6). Each cluster is different by the timing and amount of showing looking-around behavior. Cluster 1 and 2 shows participants only fixated on the front until the end of the trial. In Cluster 1, participants started at 0 degree and turned toward right (< 55 degree) and then turned toward left (> -55); yet the range was within the maximum field of view. In Cluster 2, participants shifted the head orientation only in the range of -55 degree to 0 degree. In Cluster 3, participants fixated their head toward the opposite direction fixated toward the opposite side until the end of the trial (-200 ~ -150 degrees). In Clusters 4 and 5, participants continued to see the opposite direction until the car stopped. In Cluster 4, participants looked at the opposite side (> -150) at the beginning and looked toward the AV when it approached (between 6-8 seconds, >150 degree).

In Cluster 5, participants looked toward the AV when heard the sound of AV and spotted the appearance (>200 degree). In Cluster 6, pedestrians 1) looked toward the direction of an AV approaching (> 250 degree), 2) spotted the vehicle, 3) looked towards the opposite side(<-150 degree), 4) and then looked toward the AV(> 250 degree). Comparison of count of each cluster by conditions are illustrated in Figure 7.

Cluster 1 was found overwhelmingly in No-yield condition. Interestingly, cluster 2 was not present in No-yield condition. Cluster 3 was found relatively higher in slow-yield than other conditions. Cluster 4, was found the most in No-yield condition. Cluster 5 was found the most in Fast-yield condition. Participants showed a moderate amount of looking around behavior (cluster 6) in all conditions. This suggest that 1) participants mainly fixated their head orientation up front (Cluster 1 and 2), 2) participants showed different looking-around behavioral strategy by the different action of AV.

### 4.3 Post-experiment Interview

We collected ratings on each variable: Realism, Behavior Similarity, Effects of AV's Size and AV's Yielding Behavior. Participants sat down next to the experimenter and verbally reported the response. Participants verbally report additional comments as well. We then calculated the average and standard deviation of each rating. Table 4 shows pedestrians' mean ratings for each item. Table 3 shows pedestrians' mean ratings for each item. Participants provided additional comments about the overall experience of crossing on VR crosswalk.

| Variables | M(SD) |
|---|---|
| Simulation Realism | 3.98(0.80) |
| Realistic Behavior | 4.03(0.81) |
| Yielding Behavior of AV | 4.05(0.81) |
| Size of AV | 3.40(0.68) |

Table 4. Ratio of Clusters in AV Behavior by AV Size Conditions: Simulation Realism rating is on 5-point scale (1 = "Not very realistic", 5 = "Very realistic"). Realistic Behavior rating is on 5-point scale (1="Not very similar", 5 = "Very similar"). Both ratings on Yielding Behavior of AV and Size of AV are on 5-point scale (1="Not at all", 5="Very much").

**Realism.** Most participants agreed that the virtual crosswalks felt real (3.98 out of 5, with 5 being very much realistic). Even though the participants were aware of the fact that they were in a virtual reality (a simulated setting), it did not stop them from fully focusing on the problem at hand. Participants reported that even if they knew they were in a simulated world, it did not stop them from being immersed in the situation. One participant said:

"I felt like I was a character in a 3D animated world."

One participant, however, mentioned that he got used to the environment and got bored. The experiment naturally required participants to walk and run repeatedly and as they performed the same action, this could have led to loss of interest.

**Similarity to Real-world Crossing Behavior**. Most participants reported they crossed similarly to how they usually cross, given that the crosswalk in the task were close replications of a typical alley or small road in Korea. Several participants reported that they were more cautious as they were uncertain about the AV's intention to yield. These uncertainties were mainly due to the fact that participants could not see the driver inside the AV. Some participants commented:

"I crossed as if I was crossing a real crosswalk. I felt like I had to send some signals to the driver so that I can feel safe. I waved my hands and tried to make eye contact. I acted more carefully after a crash in the beginning of the experiment."

"I couldn't see the driver in the AV, so I thought the AV wouldn't yield if it had noticed my presence."

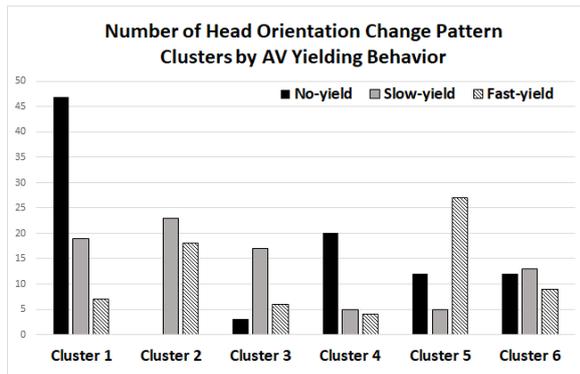

Figure 7. Number of Head Orientation Change Pattern Clusters by AV Yielding Behavior

Only few participants admitted that repetitive exposure to the similar crosswalks in the VR made them tired and less vigilant of the AV movement.

**Effect of Yielding Behavior of AV.** Most of the participants said that yielding behavior of the AV affected their crossing behavior. Participants stated that when the AV stopped quickly, they crossed rather slowly. However, when the AV stopped slowly, it took more time for them to determine whether the AV was stopping to yield and whether it was safe to cross. Some participants reported that the yielding behavior affected their judgment on safety rather than the speed of the AV.

**Effect of AV Size.** Most participants responded that the size of the vehicle affected their crossing behavior. In particular, the larger the AV, the more carefully they crossed. However, the act of caution was divided into two types rushing ahead before approaching or waiting until the AV completely passed. Interestingly, some participants reported that they perceived smaller AVs more threatening than AVs of different sizes as they seemed to approach faster. Another group of participants reported that virtual AVs made them willing to take riskier crossing decisions (e.g., crossing without looking at the AV), except for large-sized AVs. Large size AVs were too tall for them to make eye contacts, and thus led to decreased perception of safety. Here, we add part of comments participants provided:

"I didn't find other AVs threatening except for the large truck. I was scared that the driver inside wouldn't notice me."

"I found it very overwhelming when the AV was a large truck. I had a car accident when I was a child. So I waited until all cars had fully passed."

"I didn't think the size of the AV mattered until the large truck appeared. It was scary when it seemed like it passed right in front of my nose."

## 5 Discussion

In the present study, we investigated the influence autonomous vehicle related variables such as size and yielding behavior on change of head orientation of Korean pedestrians. We used virtual reality simulation to explore the impact of target variables in order to observe the direct effect of them. To our knowledge, this is the first to explore vehicle behavior in a virtual reality, especially in the perspective of a pedestrian, as most of the previous literature was focused on the view of a driver. We also took a novel approach in analysis by using a machine learning classification, Dynamic Time Warping and K-means clustering in order to find common patterns of head orientation. To our knowledge, this is the first study 1) investigated the effect of AV's explicit movement on pedestrians' trust, 2) quantified pedestrian trust as head orientation angle change, 3) and used machine learning methods to classify pedestrian's head orientation patterns.

The main findings of our research follows. First, participants avoided making an eye-contact with the AV. This was both evident in the largest ratio of Cluster 1 among head orientation patterns and in the interview. Knowing that there is no driver inside the vehicle did not affect pedestrian crossing behavior. We found that the participants deliberately avoided eye contact in a hope that the AV could yield automatically. This is in line with the research that people use deliberate communication to send a signal to the driver that he/she is crossing (Risto et al., 2017; Dey & Terken, 2017; Guéguen et al., 2015). We confirmed this in our experiment and being in a virtual world did not affect them showing similar-to-real-world response. A possible explanation of the result may be due to the cultural factors. Previous studies found that drivers in developing countries rarely yield to pedestrians (Kadali & Vedagiri 2019). In the present study and interestingly, participants raised their hand to signal their presence. This, however, could be interpreted differently in other cultures: either showing gratitude or claiming. Misinterpretation of these non-verbal signals may put pedestrians in danger.

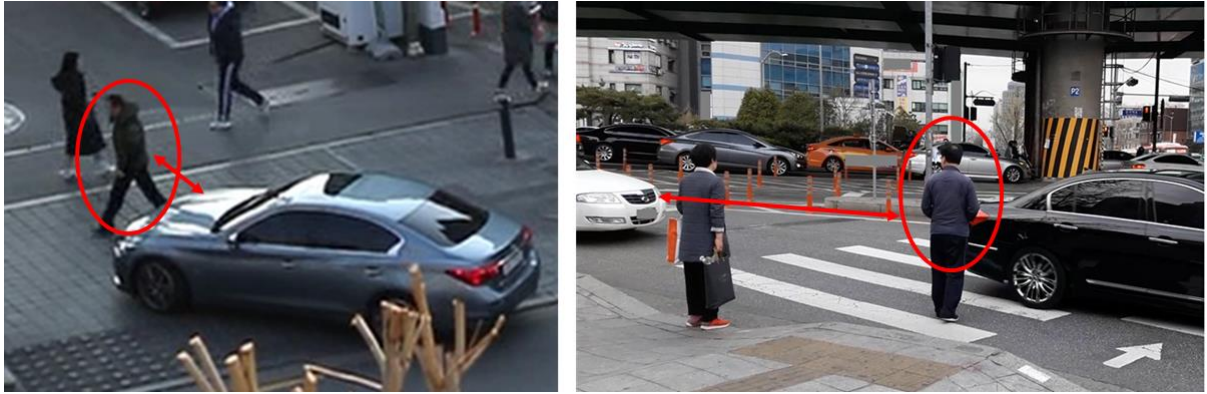

Figure 8. Example of Pedestrians Using Non-Verbal Communication to Show Intent to Cross in Korea: In the left picture, a man deliberately ignores the approaching car. The car slowly yields to the man. In the right picture, a man looks at an approaching car (white). The white car yields fast. We found people react the same to the autonomous vehicle and these behaviors were different by AV's choice to yield and size.

Second, participants chose to yield differently in reaction to AV's yielding behavior. Participants were less likely to change head orientation when AV did not yield (Cluster 1). Interestingly, slow-yield had the most variance in clusters (although still fixating-front patterns were observed most frequently). This suggests that the slower AV yields, the more confusing it is to interpret. In the interview, participants reported that they were aware of the AV approaching but were confused whether it will yield. In contrast, AV's yielding-fast behavior made participants show more of look-out behavior or look at the opposite side of the AV. A possible explanation is that fast-yielding behavior was more assuring to let them look out for other cars coming from the opposite. This, however, could be different when investigated in different contexts. Here, we used road settings that are similar to small two-way roads with no signals. Previous studies found people show different levels of trust to AV in signalized crosswalk than in unsignalized crosswalk (Jayaraman et al., 2019). In signalized crosswalk, people are more likely to accept a shorter distance between the vehicle and themselves because there is a mutual trust that both parties will not overpass the signalized lines. In an unsignalized crosswalk, the level of pedestrian trust is much lower. However, the relationship can be different depending on the culture. In Korea, it is easy to find instances of pedestrians allowing extremely short distances from vehicles (Figure 8); pedestrians easily allow short distance to the vehicle in unsignalized crosswalks. This can explain why participants mostly fixated front in our results.

Third, participants chose to yield differently in reaction to AV's size as well. Large AVs led to limited head orientation. In contrast, participants showed more head orientation in reaction to small AVs' fast yielding and not yielding. During the experiment, participants were either more likely to run as soon as they spotted the large AV or wait until the large AV completely stopped or went away. Participants reported that the large AV was most threatening; for example, one participant (180 cm, or 6 feet tall) reported that approaching large AVs looked "gigantic". We found that the size of a vehicle affects pedestrian crossing; we also confirmed the relationship between vehicle size and pedestrian behavior is not affected by knowing the absence of a driver. Therefore, the finding suggests that pedestrians will be more likely to let large AVs pass (yield) and attempt to cross in front of a small AV. This suggests that designing small AVs require more precision and discretion in technologies and algorithms that interpret pedestrian intent.

In the present study, we did not explore the effect of AV's sound on pedestrian behavior. Although we did put an engine sound on each of the beginning of trials, this was only for the purpose of alarming participants about the appearance of a vehicle from a distance. Using sounds, an AV can draw attention to the driver from distracted pedestrians (e.g., using smartphones (Horberry et al., 2019) or wearing headphones (Basch et al., 2015), or even from visually handicapped individuals (Kim et al., 2012). Future studies should investigate how AVs can practice such etiquette to appropriately use vehicle horns only in needed situations without generating too loud or discomforting noises.

## 6 Conclusion

Studying pedestrian's intent and communication is vital for both pedestrian and driver safeties. Designing an autonomous driving technology that understands pedestrians' intention and movement can reduce uncertainty on the road situation and traffic accidents. Unlike previous studies, we used virtual reality to precisely manipulate the action of AV and studied its impact on pedestrian's intent to cross and trust. 1) Pedestrians show different communication strategies in response to AV's movement, 2) Pedestrians pay less attention to the AV despite knowing it is a driverless car, 3) Yielding behavior of AV is perceived as more trustworthy, especially when yielded fast. 4) Effect of yielding behavior on pedestrian crossing behavior can differ by size of AV: large AVs reduce the need for pedestrians to communicate with the AV, whereas small AV is the opposite. 4) Pedestrian behavior in reaction to AV's behavior can differ by culture. Learning socially appropriate behaviors (i.e., etiquette) can make AVs perceived more trustful, generate more safe interactions, and reduce collisions that occur due to misinterpretation of each other's signs.

## 7 Acknowledgments

This research was supported by the *2018 Interdisciplinary Research Project* funded by the Korean Psychological Association.